\documentclass[aps, twocolumn, showpacs,preprintnumbers,amsmath,amssymb,superscriptaddress]{revtex4-2}
\usepackage{amsmath,amsfonts,amssymb,graphics,graphicx,epsfig,color,times,indentfirst,layout,hyperref,multirow,bm}
\usepackage{hyperref, soul}
\hypersetup{linktocpage,colorlinks=true}
\usepackage{threeparttable}
\usepackage{dcolumn}
\usepackage{bm}
\usepackage{color}
\usepackage{footnotebackref}

\begin{document}
\title{Nonlinear photoconductivities and quantum geometry of chiral multifold fermions}
\author{Hsiu-Chuan Hsu}\email{hcjhsu@nccu.edu.tw}
\affiliation{Graduate Institute of Applied Physics, National Chengchi University, Taipei 11605, Taiwan}
\affiliation{Department of Computer Science, National Chengchi University, Taipei 11605, Taiwan}
\author{Jhih-Shih You}\email{jhihshihyou@ntnu.edu.tw}
\affiliation{Department of Physics, National Taiwan Normal University,  Taipei 11677, Taiwan}
\author{Junyeong Ahn}
\email{junyeongahn@fas.harvard.edu}
\affiliation{Department of Physics, Harvard University, Cambridge, MA, USA}
\author{Guang-Yu Guo}\email{gyguo@phys.ntu.edu.tw}
\affiliation{Department of Physics, National Taiwan University,  Taipei 10617, Taiwan}
\affiliation{Physics Division, National Center for Theoretical Sciences, Taipei 10617, Taiwan}
\date{\today}
\begin{abstract}
Chiral multifold fermions are quasi-particles that appear only in chiral crystals such as transition metal silicides in the cubic
B20 structure (i.e., the CoSi family), and they may show exotic physical properties.
Here we study the injection and shift photoconductivities and also the related geometrical quantities for several types 
of chiral multifold fermions, including spin-1/2 as well as pseudospin-1 and -3/2 fermions, 
dubbed as Kramers Weyl, triple point and  Rarita-Schwinger-Weyl (RSW) fermions, respectively.
We utilize the minimal symmorphic model to describe the triple point fermions (TPF).
We also consider the more realistic model Hamiltonian for the CoSi family including both linear and quadratic terms.
We find that {injection currents due to circularly polarized light} are quantized as a result of the Chern numbers carried by the multifold fermions
within the linear models.
Surprisingly, we discover that in the TPF model, {the linear shift conductivities, responsible for the shift current generation by linearly polarized light,} are proportional to the pseudo spin-orbit coupling 
and independent of photon frequency.
In contrast, for the RSW and Kramer Weyl fermions, the linear shift conductivity is linearly proportional to photon frequency.
The numerical results agree with the power-counting analysis for quadratic Hamiltonians.
The frequency independence of the linear shift conductivity could be attributed to the strong resonant symplectic Christoffel symbols of the flat bands.
Moreover, the calculated symplectic Christoffel symbols show significant peaks at the nodes, suggesting that the shift currents are due to 
the strong geometrical response near the topological nodes. 
\end{abstract}
\maketitle
\section{Introduction}
Multifold fermions are types of quasi-particles that only appear in solids with particular crystal symmetries \cite{Bradlyn2016,Hasan2021}. Their pseudospin degrees of freedom are the degeneracies at the high-symmetry points in the Brillouin zone. There is no counterpart in the elementary particle model. Thus, the study of the physical properties and genuine signatures of multifold fermions in solids is of great interest. 

Recent advances in solid state physics show that the topological and geometrical properties of quantum states manifest in several physical quantities, one of which is photovoltaic effect. It is the generation of d.c. current in a noncentrosymmetric solid under the irradiation of light without an external bias. Thus, the photovoltaic effect plays an important role in the search for green energy supplications \cite{Nagaosa2017,Cook2017}. 
The photovoltaic response functions are closely related to the quantum geometrical quantities, such as connections, quantum metric and Berry curvature. 
The quantum geometric properties are related to transport in semiclassical picture. For anomalous Hall effect, Berry curvature gives rise to the anomalous velocity of carriers \cite{Xiao2010,Nagaosa2010}.  More recently, the second-order response of electrons to electromagnetic fields is shown to relate to the quantum metric and Christoffel symbols, which give rise to the gravity in momentum space \cite{Ahn2020,Smith2022}. The possibility of the quantization of quantum metric in topological semimetals has been discussed \cite{Lin2021,Lin2022}. 
In another perspective, the photoelectric response can be utilized to probe quantum geometry of Bloch states \cite{Nagaosa2017,Ahn2022,Hwang2021,Chaudhary2022}. 
Therefore, the investigation of the seemingly pure mathematical structure would deepen the theoretical and experimental understanding of solids.

The photovoltaic effect in topological semimetals have been widely studied \cite{deJuan2017,Chan2017,Chang2017, Patankar2018,Flicker2018,Ma2019,Ahn2020,Ni2020,Ipsita2020,Ni2021,Sadhukhan2021,Sadhukhan2021Role,Sekh2022}. It has been found that the Weyl semimetal possesses low frequency divergence which makes it a promising candidate for terahertz photodetectors \cite{Chan2017,Patankar2018,Ahn2020,Ni2020,Ni2021}. 
	However, for chiral symmetric Weyl semimetals, the photovoltaic response of the topological node and antinode cancels out unless the Weyl nodes are tilted \cite{Chan2017}. 
		In contrast, for chiral crystals, the Weyl points are separated in energy, as a result of the chiral symmetry breaking. There is an available energy window for nonvanishing photocurrent even for upright cones. Therefore, 
the chiral Weyl semimetals are promising materials for realizing strong photovoltaic response.   
	
The relation between the second-order photoconductivity tensors and topology has been investigated by several authors. It has been theoretically shown that in chiral symmetry broken Weyl semimetals, the circular photogalvanic response is quantized due to the Chern number of the Weyl node near the Fermi level \cite{Cook2017}. Moreover, the second-order photoconductivity is related to the connection and curvature, reflecting the geometry of Bloch states involved in the transition \cite{Ahn2022}.  

The photoconductivities in chiral multifold fermions have been studied in real materials, especially in the CoSi family of space group 198 \cite{Li2019,Ni2020,Rees2020,Xu2020,Chang2020,Sun2020,Rees2021,Ni2021,Lu2022}. The material hosts several types of topological semimetals, including, type-I, type-II Weyl semimetal and chiral multifold fermions \cite{Li2019,Ni2020,Rees2020,Xu2020,Ni2021,Lu2022,Hsieh2022}. Thus, it is a very suitable material for investigation of physical properties of topological semimetals. 

For second-order photoconducitvites, 
{several mechanisms that contribute to the second-order conductivities have been proposed, such as anomalous \cite{Rostami2018,Bhalla2022}, resonant photogalvanic \cite{Bhalla2020}, double resonance and higher-order pole \cite{Bhalla2022}. }
{In this paper, we study} two contributions, the injection and shift current. 
The injection current is related to Berry curvature \cite{deJuan2017,Ahn2020} and quantum metric \cite{Lin2021,Smith2022}, while the shift current is related to Hermitian connections \cite{Morimoto2016,Nagaosa2017,Ahn2022}. 
However, an understanding of shift current and its geometrical origin for multifold fermions have been lacking.  How momentum space quantum geometry contributes to optical response via Christoffel symbols has not been carefully examined. 
This paper aims at shedding light on this topic. 
Two model Hamiltonians for multifold fermions are studied in this paper. The first is a pseudospin-1 excitation, which is dubbed as triple point fermion (TPF). The minimal symmorphic model for TPF, of which the degenerate nodal point is protected by $C_4$ and an anti-commuting mirror symmetry,  is used in this study.
The second is the low-energy effective Hamiltoinan for space group 198. When spin-orbit coupling is switched off, the effective Hamiltoinian represents two degenerate TPFs (DTPF). In contrast, when spin-orbit coupling is included, the degenerate TPFs split into two sets of degenerate points, a spin-$3/2$ excitation, dubbed as Rarita-Schwinger-Weyl (RSW) or a four-fold fermion, and a spin-$1/2$ Weyl point. 

In this paper, we give analytical expressions of the second-order photoconductivities in terms of geometrical quantities and report the numerical results for TPF, DTPF, RSW and Kramer Weyl fermions.
The injection conductivity is shown to be related to quantum geometric tensors. The shift conductivity is not only contributed by Christoffel symbols, but also the contorsion tensors. The numerical results show that the shift conductivity can be merely given by the contorsion tensors, whereas the corresponding Christoffel symbols vanish. Our findings disclose the significance of contorsion tensors which have been overlooked in previous studies \cite{Bhalla2022}.  Moreover, for chiral fermions described by the quadratic Hamiltonian, our results show that the lowest order of the second-order photoconductivity scales as $\omega^0$. Particularly, the lowest order of the shift conductivity is proportional to the pseudo spin-orbit coupling. In contrast, the lowest order of the injection conductivity is independent of model parameters, in agreement with the quantization of circular injection conductivity.
The remainder of this paper is organized as follows. In Sec. \ref{sec:conduct}, the second-order photoconductivities and their relations to the quantum geometrical quantities are given. In Sec. \ref{sec:model}, the model Hamiltonians and the power counting analysis of the second-order photoconductivities for quadratic Hamiltonians are presented. The numerical results and discussions are given in Sec. \ref{sec:results}. Finally, the conclusion is given in Sec. \ref{sec:concl}.
\section{Second-order Photoconductivities and Quantum Geometry}\label{sec:conduct}
In this study, we consider two contributions to the d.c. response of the second-order photoconductivies \cite{Sipe2000}. 
According to their mechanisms, they are characterized into two processes, injection and shift current. The injection (shift) refers to the change of group velocity (position) during the interband transition. The topological and geometrical aspects have been discussed in literature, some of them will be reviewed in this section. 

The shift photoconductivity is given by \cite{Aversa1995,Ahn2020}
\begin{eqnarray}
	\sigma^{c,ab}_{\text{shift}}=\frac{-\pi e^3}{\hbar^2}\int\frac{d^dk}{(2\pi)^d} \sum_{n,m}f_{nm}I^{c,ab}_{mn} \delta(\omega_{mn}-\omega)
\label{eq:shiftcond}
\end{eqnarray}
where 
$\hbar\omega_{mn}=E_m-E_n$ is the energy difference between two bands, $d$ is the spatial dimension, $f_{nm}=f_n-f_m$, where $f_{n,m}$  is the Fermi-Dirac distribution. The electron charge is $-e$ and $e>0$. 
The integrand for shift conductivity is 
\begin{eqnarray}
I_{mn}^{c,ab}&=&(R_{mn}^{c,a}-R_{nm}^{c,b})r^b_{nm}r^a_{mn},
\label{eq:shiftintgrand}
\end{eqnarray}
where $R_{mn}^{c,a}$ is the shift vector
\begin{eqnarray}
R_{mn}^{c,a}&=&r_{mm}^c-r_{nn}^c+i\partial_c {\rm log\ } r_{mn}^a
\label{eq:shiftvector}
\end{eqnarray}
and $r_{mn}^a=\langle m| i\partial_a|n\rangle$ is the Berry connection \cite{foot:logr}. 
The term  $r^b_{nm}r^a_{mn}$ is the real part of the band-resolved quantum geometric tensor, defined as $Q^{ba}=\sum_{n\in\text{occ}}\sum_{m\in\text{unocc}}r_{nm}^br_{mn}^a$ \cite{Provost1980,Tan2019}, where (un)occ denotes the (un)occupied bands. The real part of $Q^{ba}$ is the quantum metric $g^{ba}$, while the imaginary part is proportional to Berry curvature $\Omega^{ba}$. The relation is 
\begin{eqnarray}
Q^{ba}=g^{ba}-\frac{i}{2}\Omega^{ba}.
\label{eq:qgt}
\end{eqnarray} 
Eq. \ref{eq:shiftintgrand} can also be written as $i(r^b_{nm}r^a_{mn,c}-r_{nm,c}^br_{mn}^a)$, where $r_{mn,c}^a=\partial_cr_{mn}^a-i(r_{mm}^c-r_{nn}^c)r_{mn}^a$. 
Notably, $r_{nm}^br_{mn,c}^a$ is a geometrical quantity for the quantum states \cite{Ahn2022}. 
We define $C_{nm}^{bca}=r_{nm}^br_{mn,c}^a$. 
The non-abelian Berry connections form tangent vectors in the manifold of the Bloch states. In the subpace of the tangent vectors, $C_{nm}^{bca}$ is the Hermitian connection that defines the covariant derivative. Note that the order of the index for Hermitian connections is $bca$ for the conductivity tensor $cab$.
{
$C_{nm}^{bca}$ is in general complex. The real part of $C_{nm}^{bca}$ is the metric connection and the negative imaginary part of $C_{nm}^{bca}$ is the symplectic connection.}
Note that the metric connection here is different from the Levi-Civita connection
\begin{eqnarray}
	\Gamma^{bca}_{nm} =
	\frac{1}{2}\left(
	\partial_cg_{nm}^{ba}+\partial_a g_{nm}^{bc} -\partial_b g_{nm}^{ca}
	\right)
\end{eqnarray}
when the number of bands in the system exceeds two. The difference is characterized by the contorsion tensors. We define a generalized complex-valued contorsion tensor $K^{bca}_{nm}$ such that it satisfies
\begin{eqnarray}\label{eq:chris}
\Gamma_{nm}^{bca}&=&\text{Re}\left[ C_{nm}^{bca}-K_{nm}^{bca}\right]
\end{eqnarray}
and define the corresponding symplectic part by
\begin{eqnarray}\label{eq:sympchris}
	\tilde{\Gamma}_{nm}^{bca}&=&-\text{Im}\left[ C_{nm}^{bca}-K_{nm}^{bca}\right].
\end{eqnarray}
The expression of the contorsion tensor is given in Appendix A. The fully symmetric part with respect to the permutation of $b,c$, and $a$ of the $\text{Im}\left[K_{nm}^{bca}\right]$ is chosen to be zero. 
Eq. \ref{eq:chris} (Eq.\ref{eq:sympchris}) is the Levi-Civita connection part of the metric (symplectic) connection.  We refer to $\Gamma^{bca}_{nm}$ ($\tilde{\Gamma}^{bca}_{nm}$) as (symplectic) Christoffel symbols in this paper. 


{The real part of the shift photoconductivity (called linear shift photoconductivity hereafter, as in \cite{Ahn2020}), which is responsible for the shift current generation by linearly polarized light,} can be written in terms of the symplectic Christoffel symbols,
\begin{widetext}
\begin{eqnarray}
	{\sigma^{c,ab}_{\text{shift;L}}}=\frac{-\pi e^3}{\hbar^2}\int\frac{d^dk}{(2\pi)^d} \sum_{n,m}f_{nm}\left(
		\tilde{\Gamma}_{nm}^{bca}+	\tilde{\Gamma}_{nm}^{acb}
		-\text{Im}\left[
		K_{nm}^{bca} +	K_{nm}^{acb}
		\right]
	\right)
	\delta(\omega_{mn}-\omega).
	\label{eq:linshch}
\end{eqnarray}
{The imaginary part of the shift photoconductivity (called circular shift photoconductivity hereafter), which is the response to circularly polarized light,} can be written in terms of the Christoffel symbols of the first kind, 
\begin{eqnarray}
	{\sigma^{c,ab}_{\text{shift;C}}}=\frac{-\pi e^3}{\hbar^2}\int\frac{d^dk}{(2\pi)^d} \sum_{n,m}f_{nm}\left(
	{\Gamma}_{nm}^{bca}-{\Gamma}_{nm}^{acb}
	-\text{Re}\left[
	K_{nm}^{bca} -	K_{nm}^{acb}
	\right]
	\right)
	\delta(\omega_{mn}-\omega).
	\label{eq:circshch}
\end{eqnarray}
\end{widetext}
For numerical calculations, $C_{nm}^{bca}$ is written in terms of the velocity operators and double derivatives of the Hamiltonian
\begin{eqnarray}
	C_{nm}^{bca}&=&\frac{v_{nm}^b}{\omega_{mn}^2}
	\bigg[
	w_{mn}^{ac}-\frac{v_{mn}^c\Delta_{mn}^a+v_{mn}^a\Delta_{mn}^c}{\omega_{mn}}\nonumber\\
	&+&\sum_{p\neq m,n}
	\left(
	\frac{v_{mp}^c v_{pn}^a}{\omega_{mp}}-
	\frac{v_{mp}^a v_{pn}^c}{\omega_{pn}}
	\right)
	\bigg],
\label{eq:shiftnum}
\end{eqnarray} 
where $w_{mn}^{ac}=\hbar^{-1}\langle m|\frac{\partial^2H}{\partial k_a\partial k_c}|n\rangle$, 
$v_{mn}^a=\hbar^{-1}\langle m|\frac{\partial H}{\partial k_a}|n\rangle$, 
$\Delta_{mn}^a=v_{mm}^a-v_{nn}^a$.  

The injection conductivity is given by
\begin{eqnarray}
\sigma^{c,ab}_{\text{inj}}=-\tau\frac{2\pi e^3}{\hbar^2}\int\frac{d^dk}{(2\pi)^d} \sum_{nm}f_{nm}\Delta_{mn}^c r_{nm}^br_{mn}^a
 \delta(\omega_{mn}-\omega), \nonumber\\
\label{eq:injcond}
\end{eqnarray}
where $\tau$ is the relaxation time \cite{foot:relaxapp}.
For topological semimetal that carries topological charges under the irradiation of circular polarized light, trace of the injection conductivity is quantized, dubbed as quantized circular photogalvanic effect (CPGE) \cite{deJuan2017, Flicker2018}.  $\sum_{\text{cycl}}\sigma^{c,ab}=i\beta_{0}C\tau$, where $\sum_{\text{cycl}}$ denotes the summation over $c,a,b$ in cyclic permutation, $C$ is the topological charge of the semimetal, $\beta_0=\frac{\pi e^3}{h^2}$ and $h$ is the Planck constant. 

  The explicit forms of the injection conductivity tensors in terms of quantum geometrical tensor are given below.
  {By taking the real part of Eq. (\ref{eq:injcond}),} 
 the linear injection photoconductivity is 
 \begin{eqnarray}
 {\sigma^{c,ab}_{\text{inj;L}}}=-\tau\frac{2\pi e^3}{\hbar^2}\int\frac{d^dk}{(2\pi)^d} \sum_{n,m}f_{nm}
 \Delta_{mn}^c g^{ba}_{nm}
 \delta(\omega_{mn}-\omega).\nonumber\\
 \label{eq:injmet}
 \end{eqnarray}
{By taking the imaginary part of Eq. (\ref{eq:injcond}),} 
 the circular injection photoconductivity, is  
 \begin{eqnarray}
 {\sigma^{c,ab}_{\text{inj;C}}}=\tau\frac{\pi e^3}{\hbar^2}\int\frac{d^dk}{(2\pi)^d} \sum_{n,m}f_{nm}
 \Delta_{mn}^c \Omega^{ba}_{nm}
 \delta(\omega_{mn}-\omega),\nonumber\\
 \label{eq:injberr}
 \end{eqnarray}
where $g^{ba}_{nm}={\rm {Re}}\left[r^b_{n}r^a_m\right]$ and $\Omega^{ba}_{nm}=-2{\rm {Im}}\left[r^b_{n}r^a_m\right]$ are band resolved quantum metric and Berry curvature, respectively.

 In the numerical calculation, the Dirac delta function in the equations is replaced with the Lorentzian function 
 \begin{eqnarray}
 \mathcal{L}=\frac{1}{\pi}\frac{\gamma/2}{(\omega_{mn}-\omega)^2+(\gamma/2)^2}, 
 \end{eqnarray}
 where $\gamma$ is the broadening.  

\section{model Hamiltonians and power counting analysis}\label{sec:model}
The model Hamiltonians of the triple point fermion and the multifold fermions in the CoSi family are introduced in this section. 
 
The first model Hamiltonian considered in this paper is the minimal symmorphic model for TPF \cite{Fulga2017}. 
This model can be viewed as stacked layers of Chern insulators along the $z$-direction and thus the time-reversal symmetry is broken.
The topological charge of the Weyl point is $\pm 2$. The band dispersion is a result of the coupling between the quadratic Weyl point and a additional flat band via pseudo spin-orbit coupling. 
The {minimal two-band} Hamiltonian for the quadratic Weyl fermions is
\begin{eqnarray}
H_{q}(\vec{k})&=&\left[s(2-\cos(k_x)-\cos(k_y))-2t\cos(k_z)
\right]\sigma_z \nonumber\\
&+&2b\sin(k_x)\sin(k_y)\sigma_y\nonumber\\
&+&2b\left[\cos(k_x)-\cos(k_y)
\right]\sigma_x,
\end{eqnarray}
where $b$ is the pseudo spin-orbit coupling strength, $s$ is the on-site hopping strength. 
The $z$-direction hopping term $t$, lattice constant $a$ and $\hbar$ are taken to be $1$ in this model. 
The Weyl points are at $(0,0,\pm\pi/2)$ and of opposite chirality.
By introducing a flat band that couples to the quadratic Weyl fermions, we obtain an effective {$3\times 3$} Hamiltonian for the triple point fermion \cite{Fulga2017,Hsu2022Disorder}
\begin{eqnarray}
H_t(\vec{k})=\begin{pmatrix}
\multicolumn{2}{c}{\multirow{2}{*}{$H_q$}}& \lambda_+^{\dagger}\\
&&\lambda_-^{\dagger}\\
\lambda_+& \lambda_-&0
\end{pmatrix},
\label{eq:tpfham}
\end{eqnarray}
where $\lambda_{\pm}=\lambda e^{i(\phi\pm \pi/4)}(\sin k_x\mp i \sin k_y)$. 
Hereafter, we choose $\lambda=\sqrt{2}$ and $\phi=\pi/2$ for isotropic dispersion (to the lowest order). 

\begin{table}[]
\caption{The sign change for the matrix elements under $z$-mirror symmetry ($M_z$).}
	\renewcommand{\arraystretch}{1.3}
\begin{tabular}{|lllllll|}
	\hline
	\multicolumn{7}{|c|}{$O_{mn}(k)\Rightarrow \pm O_{mn}(k)$}                                                                                                                                                                                                           \\ \hline
	\multicolumn{1}{|l|}{$r_{mn,z}^z(k)$} & \multicolumn{1}{l|}{$r_{mn,{c\neq z}}^z(k)$} & \multicolumn{1}{l|}{$r_{mn,{z}}^{a\neq z}(k)$} & \multicolumn{1}{l|}{$r_{mn}^z(k)$} & \multicolumn{1}{l|}{$v_{mn}^z(k)$} & \multicolumn{1}{l|}{$r_{mn}^{a\neq z}(k)$} & $v_{mn}^{a\neq z}(k)$ \\ \hline
	\multicolumn{1}{|l|}{+1}              & \multicolumn{1}{l|}{-1}                      & \multicolumn{1}{l|}{-1}                        & \multicolumn{1}{l|}{-1}            & \multicolumn{1}{l|}{-1}            & \multicolumn{1}{l|}{+1}            & +1            \\ \hline
\end{tabular}
\label{tab:mirror}
\end{table}

\begin{figure}[]
        \includegraphics[width=0.48\textwidth]{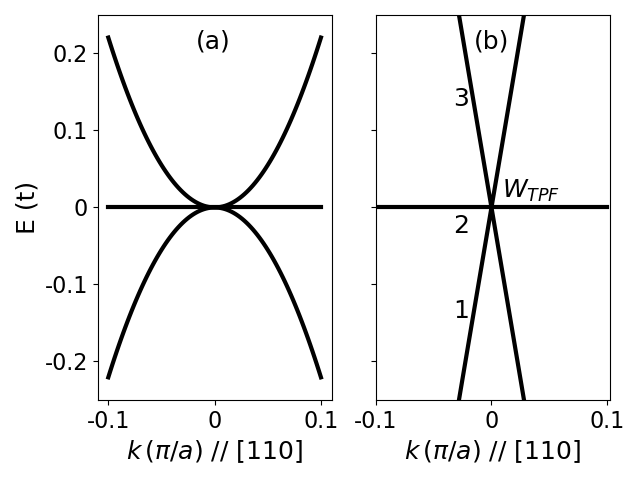}
        \caption{The energy band along [110] direction for $H^{t}$ [Eq. (6)] to the quadratic order with $b=1, s=1$. (a) $\lambda=0$. (b) $\lambda=\sqrt{2}$. The numbers annotated on the figure labels the band indexed from low to high energy. The energy at which the TPF lies is denoted by $W_{TPF}$.
        }
        \label{fig:tpfdisp}
\end{figure}

The coupling between the flat band and $H_q$ preseves the symmetry of $H_q$. 
Both Hamiltonians obey $C_4$ rotation symmetry and anticommute with mirror symmetry $R_{xy}$ that maps $x\leftrightarrow y$, 
preserving chiral symmetry, while time-reversal symmetry is broken. The two opposite topological nodes are related by the mirror symmetry along $z$-direction $M_z$. The sign change of the matrix elements for the photoconductivities under $M_z$ are shown in Table \ref{tab:mirror}. The conductivity tensor of which components with odd numbers of z changes sign for opposite nodes, leading to vanishing response for the lattice. To break the chiral symmetry, an additional term that breaks the mirror symmetry along $z$- direction, $d\sin(k_z)I_{3\times3}$, is added to the Hamiltonian (Eq. \ref{eq:tpfham}), where $I_{3\times3}$ is the $3\times 3$ identity matrix \cite{deJuan2017}. Thus, the chiral symmetry is broken and the two TPFs are separated in energy. 
In the following, we consider the response near one of the nodes. 
We consider the low-energy expansion of the Hamiltonian Eq. (\ref{eq:tpfham}) up to quadratic order of $k$ near the node. Thus, $\lambda_{\pm}=\lambda e^{i(\phi\pm \pi/4)}(k_x\mp ik_y)$ and $H_q$ becomes
\begin{eqnarray}
H_{q}(\vec{k})=\left[s\frac{k_x^2+k_y^2}{2}+2ck_z
\right]\sigma_z +\nonumber\\
2 b k_xk_y\sigma_y+
b\left[k_y^2-k_x^2
\right]\sigma_x,
\label{eq:hq}
\end{eqnarray}
where $c=\mp1$ is the chirality of the Weyl point for the node at $(0,0,\pm\pi/2)$. 
The quadratic term in the diagonal does not change the Chern number of the bands. Thus, changing the value of $s$ can be treated as a smooth deformation to the Hamiltonian.
The eigenenergies are $0,$ and
\begin{eqnarray}
	\pm\frac{1}{2}\sqrt{16k^2+(4b^2+s^2)k_{\rho}^4+8sk_zk_{\rho}^2}
\end{eqnarray}
where $k^2=k_x^2+k_y^2+k_z^2, k_{\rho}^2=k_x^2+k_y^2$.
The dispersion relations for $\lambda=0, \sqrt{2}$ with $b=1,s=1$ are shown in Fig. \ref{fig:tpfdisp}. For $\lambda=0$, the upper and lower bands are quadratic, while for  $\lambda=\sqrt{2}$, the upper and lower bands disperse linearly. The spin-excitation sits at zero energy, labeled by $W_{TPF}$. 

For a more realistic model, we take the effective Hamiltonian for transition metal silicides that belong to the space group 198. There are one threefold rotation symmetry along (111) axis and three twofold screw symmetries along the $x, y$ and $z$ axis for this group \cite{Tang2017,Chang2017}. 

\begin{figure}[]
\includegraphics[width=0.48\textwidth]{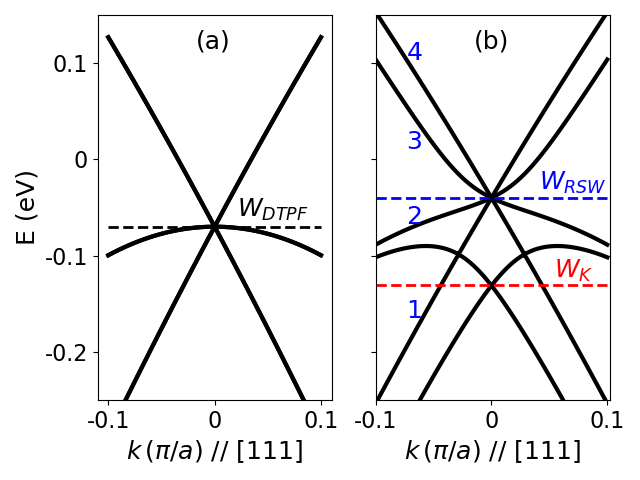}
\caption{Energy bands along [111] direction for $H_{\Gamma198}$ [Eq. (\ref{eq:hg198})] to the quadratic order.
(a) Without spin-orbit coupling. The bands are doubly degenerate. The black dashed line denotes the energy level
at the double TPF node. (b) With spin-orbit coupling. The blue (red)dashed line indicates the energy levels of
the RSW node (Kramer Weyl). The blue numbers denote the band index of the RSW node. The zero energy denotes the Fermi level.
        }
\label{fig:198disp}
\end{figure}

In order to isolate the multifold fermions at high symmetry point, we expand the tight-binding Hamiltonian to the second order of crystal momentum $k$.  
The effective low-energy Hamiltonian for $\Gamma$ point is \cite{Chang2017}
\begin{eqnarray}\label{eq:hg198}
	H_{\Gamma198}&=&\sum_{i }(H^{(i)}_o+H^{(i)}_{SOC}), 
\end{eqnarray}
where $H^{(1,2)}_o$ is the spinless part, $H^{(1,2)}_{SOC}$ is the spin-orbit coupled term and the superscripts $(1,2)$ denote the order  in momentum $k$ of the expansion. For the effective Hamiltonian to the linear order, $i=1$. For the quadratic order, the summation runs over $i=1,2$. 
Each part of the Hamiltonian is given by 
\begin{widetext}
\begin{eqnarray}
	H^{(1)}_o&=&3v_2+v_1\left[\tau_x+\tau_x\mu_x+\mu_x
	\right]
	+\frac{v_p}{2}\left[\mu_yk_x+\tau_y\mu_zk_y+\tau_y\mu_xk_z
	\right]\\
	H^{(1)}_{SOC}&=&v_r\left[\tau_y\sigma_z+\tau_x\mu_y\sigma_x+\tau_z\mu_y\sigma_y
  	\right]
  	+
  	\frac{v_s}{2}\left[
  	\tau_x\sigma_xk_x+\tau_x\mu_x\sigma_yk_y+\mu_x\sigma_zk_z
  	\right]\\
  	H_o^{(2)}&=&\frac{-v_2 k^2}{2}+\frac{-v_1}{8}\left[\tau_x(k_x^2+k_y^2)+\tau_x\mu_x(k_y^2+k_z^2)+\mu_x(k_z^2+k_x^2)
  	\right]
  	\\
  	H_{SOC}^{(2)}&=&\frac{-v_r}{8}\left[\tau_y\sigma_z(k_x^2+k_y^2)+\tau_x\mu_y\sigma_x(k_y^2+k_z^2)+\tau_z\mu_y\sigma_y(k_x^2+k_z^2)
  	\right]\nonumber\\
  	&+&
  	\frac{v'_r}{4}\left[
  	\tau_y\mu_z\sigma_xk_xk_y+\tau_y\mu_x\sigma_yk_yk_z+\mu_y\sigma_zk_zk_x
  	\right],
\end{eqnarray}
\end{widetext}
where $\tau, \mu, \sigma$ are Pauli matrices and lattice constant $a$ has taken to be $1$. 
 The parameters are obtained from fitting to the first-principle calculations. For RhSi, the fitted parameters are $v_1=0.55, v_2=0.16, v_p=-0.76, v_r=-0.03, v'_r=0.01, v_s=-0.04$ (eV) \cite{Chang2017}. The tight-binding model preserves the screw and threefold rotation symmetry of the space group 198. It was constructed with symmetry-allowed nearest neighbor hoppings \cite{Chang2017}.

When the spin-orbit coupling is turned off, i.e. $v_r, v'_r, v_s=0$ ,  there are two degenerate spin-1 excitation at $\Gamma$ point in the Brillouin zone. The energy band diagram for the quadratic Hamiltonian without spin-orbit coupling is shown in Fig. \ref{fig:198disp}(a). The bands show spin-1 excitation and are doubly degenerate, dubbed as double TPF. The node locates at energy $W_{DTPF}=-0.07$ eV.  
The low energy dispersion is similar to that of $H_t$, although with different symmetry properties from $H_{t}$. Therefore, the two models have different nonvanishing components of the optical conductivities even though the pseudospin degrees of freedom are the same. 

When SOC is turned on, the six-fold degenerate point splits up into two sets \cite{Tang2017}, as denoted by dashed lines in the band diagram in Fig. \ref{fig:198disp} (b). One is the fourfold degenerate point which is a pseudospin -$3/2$ excitation and named as RSW fermion. The other is the twofold crossing point which is a spin-$1/2$ excitation. The energy of each node is $W_{RSW}=-0.04$ eV for RSW and $W_K=-0.131$ eV for Kramer Weyl. The energy levels for the three nodes studied in this model are listed in table \ref{tab:energy}. Because the degenerate point is at $\Gamma$ point, which is one of the time-reversal invariant momentum, the two-fold degenerate point is called a Kramer Weyl \cite{Chang2018topological}. The effective Hamiltonian for the Kramer Weyl is
\begin{eqnarray}
H_{K}=\vec{k}\cdot\vec{\sigma},
\end{eqnarray}
where $\mathcal{\sigma}$ is the Pauli matrix for electron spin, not pseudospin degrees of freedom. As a result, 
the real spin of a Kramer Weyl align along the principal axis $k_x,k_y,k_z$ \cite{Chang2018topological}. 

\begin{table}[]
\caption{Notations for energy levels at each topological node for $H_{\Gamma 198}$.}
\begin{tabular}{|l|l|}
\hline
$W_{DTPF}$ & -0.07 eV  \\ \hline
$W_K$      & -0.131 eV \\ \hline
$W_{RSW}$  & -0.04 eV  \\ \hline
\end{tabular}
\label{tab:energy}
\end{table}

 {\it Power counting analysis.}
The resonance effect of photo response in topological semimetals is interesting, because it suggests the potential application as terahertz photodetectors. By dimension analysis, the dependence of the shift and injection conductivity on photon frequency can be revealed. In previous studies \cite{Yang2018,Ahn2020}, the analysis was constrained for k-linear Hamiltonian. Since in our study, the quadratic terms have significant roles, we will include linear and quadratic terms in the Hamiltonian for dimension analysis. The following analysis considers three dimensional case, i.e. $d=3$.
The dimension of the Hamiltoinan is
 \begin{eqnarray}
         H\sim \hbar v k +\hbar v' k^2 
 \end{eqnarray}
 and the eigenenergy is denoted by $E$.
 Thus, the dimension for Berry connection is
 \begin{eqnarray}
         r \sim \frac{1}{E} \frac{\partial H}{\partial k}=\frac{\hbar v +\hbar v' k}{E}. 
 \end{eqnarray}
 For $E$ in the denominator, to the lowest order of $k$ gives $E\approx \hbar v k$, Thus,
 \begin{eqnarray}
         r \sim \frac{1}{k}+\frac{v'}{v}
 \end{eqnarray}
and
 \begin{eqnarray}
        r^3&\sim \frac{1}{k^3} 
        +\frac{3}{k^2}\frac{v'}{v} 
        + \frac{3}{k}\left(\frac{v'}{v}\right)^2 
        + \left(\frac{v'}{v}\right)^3,
 \end{eqnarray}
to the lowest order $\omega\sim vk$.   

The delta function, $\delta(\omega_{mn}-\omega)$, has dimension $\omega^{-1}$. Thus, the shift conductivity scales as
 \begin{eqnarray}
        \sigma_{sh}&\sim \frac{e^3}{\hbar^2}
        \left(
         \frac{a_{-1}}{\omega} 
         +a_0\frac{v'}{v^2} +a_1\frac{v'^2}{v^4}\omega  +\text{H.O.T.}
        \right)
        \label{eq:shpower}
 \end{eqnarray}
where $a_{-1,0,1}$ are dimensionless coefficients {given by the momentum space integration in Eq. \ref{eq:shiftcond}.}
Note that the $a_{-1}\omega^{-1}$ diverging term is contributed only by the k-linear terms in the Hamiltonian and vanishes for upright Weyl cones \cite{Yang2018,Ahn2020}, which is the case for the multifold fermions considered in this study.
Thus, $a_{-1}=0$. The second term shows that the shift conductivity is independent of $\omega$, but proportional to $v'$. The similar result was found in a previous study  that shows the linear shift conductivity for Dirac surface state is linearly dependent on the warping term and independent of photon frequency \cite{Kim2017}.

For injection conductivity, $\Delta\sim v+v'k$ and $r^2\sim k^{-2}+2\frac{v'}{vk}+(\frac{v'}{v})^2$,
 \begin{eqnarray}
        \sigma_{inj}&\sim\frac{\tau e^3}{\hbar^2} 
        \left(c_0  
                               +c_1\frac{v'}{v^2} \omega +\text{H.O.T.}
                                \right),
 \end{eqnarray}
 and $c_{0,1}$ are dimensionless coefficients {given by the momentum space integration in Eq. \ref{eq:injcond}.}
 The leading term, which is independent of frequency, does not depend on the model parameters. This term corresponds to the quantization of circular injection conductivity. 
{The values of the coefficients $a_{-1,0,1}$ and $c_{0,1}$ are determined by the details of the Hamiltonian.}

\section{Numerical calculations}\label{sec:results}
In this section, we present the calculated second order photoconductivity spectra and also related geometric quantities
for the model Hamiltonians described in the preceeding section. 
\subsection{Triple point fermions}
\begin{figure}
        \includegraphics[width=0.49\textwidth]{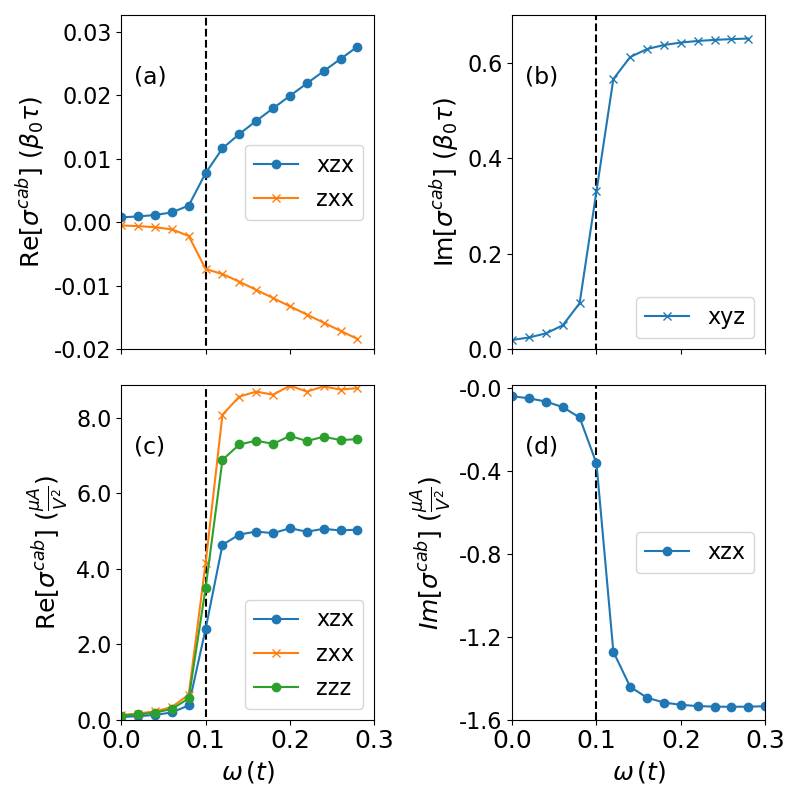}
\caption{Some of the nonvanishing components of the photoconductivity tensors for the TPF model.
(a) Linear injection, (b) circular injection, (c) linear shift and (d) circular shift conductivity.
For all the panels, the chemical potential is set to $-0.1 t$.
The vertical dashed line denotes that $\omega = |\mu|$.}
\label{fig:tpfall}
\end{figure}

The only symmetry for the effective triple point fermion model $H_t$ in the low-energy expansion is $C_4$ symmetry along the $z$ axis. 
As a result of the lowest symmetry, $H_t$ has more nonzero components of second order photoconductivity than the CoSi family.  
Furthermore, because of the broken time-reversal symmetry, all four types of the photocurrents are present, namely, 
linear and circular shift currents as well as circular and linear injection currents~\cite{Ahn2020}.
From symmetry analysis, there are 11 nonvanishing linear and 10 nonvanishing circular conductivity tensor elements \cite{Boyd}. 
Among them, there are four (three) independent linear (circular) conductivity elements.
For simplicity, we show the most prominent conductivity elements in Fig. \ref{fig:tpfall}.

\begin{figure}[]
\includegraphics[width=0.5\textwidth]{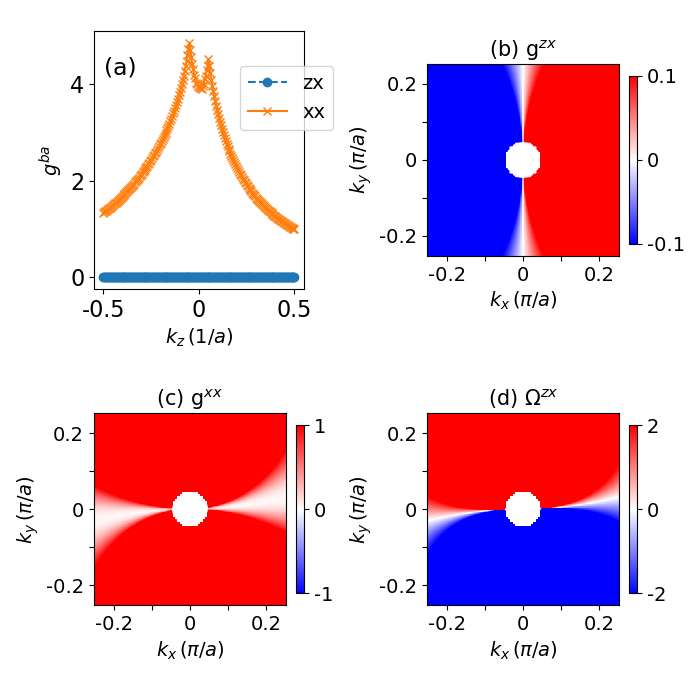}
\caption{ Quantum metric tensor elements related to the linear injection conductivity (a, b, c)
[components $xzx$ and $zxx$] and circular injection (d) conductivity [component $xyz$], respectively, for the TPF model with $\mu=-0.1t$.
(b-d) are plotted on the $k_z=0$ plane for the TPF model.  
}
\label{fig:tpfmetric}
\end{figure}

{\it Linear injection current.} The linear injection conductivity spectra are shown in Fig. \ref{fig:tpfall} (a). 
The $xzx$ and $zxx$ components are both linear with photon frequency. The linear injection conductivity 
is related to the quantum metric $g^{ba}$. In Fig. \ref{fig:tpfmetric} (a), $g^{zx}$ and $g^{xx}$ are plotted as a function of $k_z$. 
The metric element $g^{xx}$ shows a more drastic change near the node $k_z=0$, while $g^{zx}$ is zero along $k_z$. As shown in Fig. \ref{fig:tpfmetric} (b), $g^{zx}$ on the $k_z=0$ plane is an odd function in $k_x$. Therefore, the integration over the plane is zero. For linear injection conductivity ${\sigma_{inj}^{xzx}}^L$, $g^{zx}$ is multiplied by $\Delta_{mn}^x$, which is also an odd function, and the momentum space integration gives rise to nonvanishing values, as shown in Fig. \ref{fig:tpfall} (a). 
The distribution of $g^{xx}$ on the $k_z=0$ plane is shown in Fig. \ref{fig:tpfmetric} (c). 
For linear injection conductivity ${\sigma_{inj}^{zxx}}^L$, $g^{xx}$ is multiplied by $\Delta_{mn}^z$, which is a constant because the Hamiltonian is linear in $k_z$.   The values are all positive. Thus, ${\sigma_{inj}^{zxx}}^L$ is proportional to the momentum space integration of the quantum metric $g^{xx}$.

\begin{figure}[]
\includegraphics[width=0.45\textwidth]{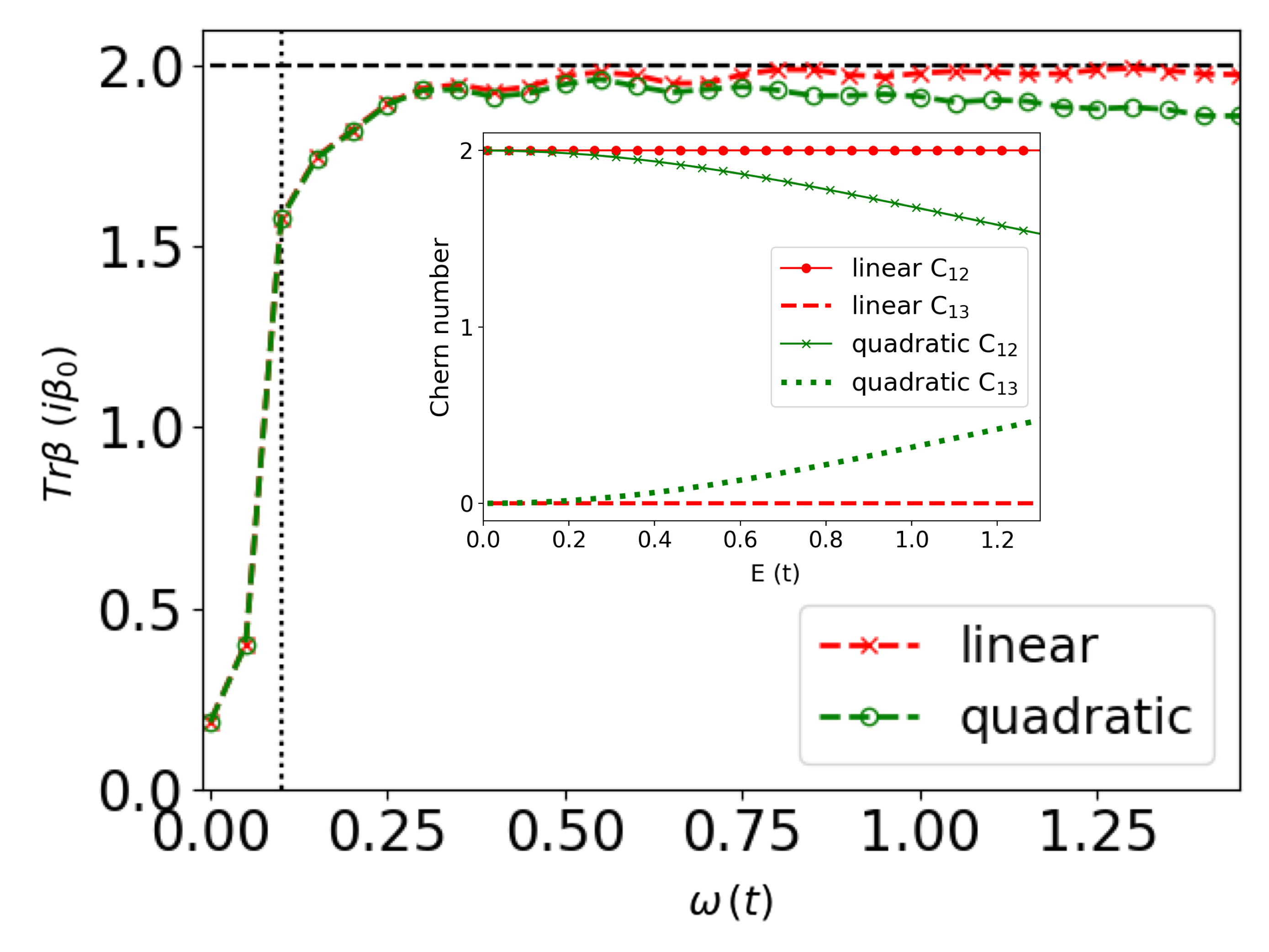}
\caption{Circular injection conductivity for linear and quadratic Hamiltonians of the TPF model. $\beta_{cc}=\sum_{a,b}\epsilon_{cab}\sigma^{cab}/\tau$, where $\epsilon_{cab}$ is the Levi-Civita symbol.
}
\label{fig:tpfcpge}
\end{figure}

{\it Circular injection current.} The circular injection conductivity is shown in Fig. \ref{fig:tpfall} (b),
which is related to the Berry curvature. The Berry curvature is an antisymmetric tensor and thus its diagonal elements $\Omega^{aa}$
vanish. Therefore, only the nondiagonal element $\Omega^{zx}$ of Berry curvature is shown in Fig. \ref{fig:tpfmetric} (d).
Clearly, $\Omega^{zx}$ is  {approximately} odd in $k_y$. Thus, when multiplied by $\Delta_{mn}^y$, the integral gives rise to 
nonvanishing circular injection current element in Fig. \ref{fig:tpfall} (b). {The $C_4$ symmetry requires that $\Omega^{zx}(-k_x,-k_y)=-\Omega^{zx}(k_x,k_y)$, but does not guarantee that $\Omega^{zx}(k_x,-k_y)=-\Omega^{zx}(k_x,k_y)$. The analysis is given in Appendix \ref{app:sym}.}
When the photon frequency is larger than the chemical potential, the value saturates at $\sim$0.65. 
This value is close to one-third of the topological charge for TPF. When taking the trace of the injection conductivity tensor, 
the value is close to the Chern number, $albeit$, with slight deviation. The deviation results from the nonzero Chern number 
between each pair of bands for the quadratic Hamiltonian.  
In Fig. \ref{fig:tpfcpge}, the $\sum_{\text{cycl}}\sigma^{c,ab}$ spectrum for $H_{\Gamma198}$ to the linear order is shown. 
Clearly, with the linear order expansion of the Hamiltonian, the conductivity is quantized at $2$, the Chern number of the Weyl node. 
While with the second-order expansion, the conductivity shifts away from the integer at higher photon frequency. This is due to the nonzero Berry curvature between a pair of bands. For injection and shift current, only the interband transitions are considered. Thus, we write the Chern number as combination of the Berry curvature between pairs of band. Assume only the lowest band is occupied and the bands are indexed from $1$ to $3$ starting from the lowest energy band. The Chern number is decomposed into 
\begin{eqnarray}
	C&=&C_{12}+C_{13},
\end{eqnarray}
where $C_{nm}$ is obtained from the the surface integration of the Berry curvature ($\Omega_{nm}(\theta, \phi)$),
\begin{eqnarray}
	C_{nm}&=&\frac{1}{2\pi}\int_0^{\pi} d\theta \int_0^{2\pi} d\phi \Omega_{nm}(\theta,\phi),\\
	\Omega_{nm}(\theta,\phi)&=&{-2}
	\ Im \frac{\langle m|\frac{\partial H}{\partial \theta}|n\rangle\langle n|\frac{\partial H}{\partial \phi}|m\rangle}
	{(E_n-E_m)^2}.
\end{eqnarray}
The calculation is done in spherical coordinate and the Chern number is obtained after integrating the Berry curvature on the constant energy surface. 
For quadratic Hamiltonian, $C_{13}$ becomes nonzero at higher energy, as shown in the inset of Fig. \ref{fig:tpfcpge}. 
As a result, the cyclic trace of the injection conductivity between the optically active pair of bands is not quantized for the quadratic Hamiltonian. 


{\it Linear shift current.} 
Fig. \ref{fig:tpfall}(c) shows the linear shift conductivities. ${zxx}$,
${xzx}$ and ${zzz}$ components are independent of photon frequency after the photon frequency is larger than chemical potential. 
The results is the lowest order in $\omega$, as suggested by Eq. \ref{eq:shpower}.
To understand the numerical results, we resort to the analytical solutions.  For analytical calculation, we use Eq. \ref{eq:shiftvector} with Berry connections. Below, the results for zzz and zxx components are presented. 
We define $I^{cab}=\sum_{nm}f_{nm}I^{cab}_{nm}$. 
To the lowest order of $k$, the analytical form of $I^{cab}_{nm}$ for the isotropic cone is, 
\begin{eqnarray}
I^{zzz}\approx
\frac{3 b \left(k_x^2+k_y^2\right)^2}
{4k^6}
\label{eq:Izzz}
\end{eqnarray} 
and
\begin{eqnarray}
I^{zxx}&\approx&
\frac{ b \left(2 k_x^4+5 k_x^2 \left(k_y^2+k_z^2\right)+3 k_y^2 \left(k_y^2+k_z^2\right)\right)}
{4k^6}.
\label{eq:Izxx}
\end{eqnarray}
It shows that the linear shift current depends linearly on the pseudo spin-orbit coupling $b$. 
After inserting the integrand to Eq. \ref{eq:shiftcond} and converting to spherical coordinate, $d^3k$ becomes $d\Omega_k k^2dk$, where $\Omega_k$ is the solid angle in k-space,
and $\delta(\omega_{mn}-\omega)$ is replaced with $\delta(k-k(\omega))/|dE/dk|$, where $|dE/dk|\sim 2$ in the linear order of $k$ and $k(\omega)=\omega/2$. 
The integral becomes 
\begin{eqnarray}
\sigma^{c,ab}=\frac{- e^3}{2 h^2}\int d\Omega_k\int k^2 dk\frac{\delta(k-k(\omega))}{2}I^{cab}.
\end{eqnarray}
One obtains
$\sigma^{zzz}\approx7.07\, b\, \mu A/V^2$  and 
$\sigma^{zxx}\approx 8.99\, b\, \mu A/V^2$. 
In Fig. \ref{fig:tpfall}, $b=1$ is used for the numerical calculation. 
The analytical values are close to the numerical values with less than $4\%$ error. Thus, the plateau corresponds to a model dependent value. 
Similarly, the $xzx$ component for both the linear and circular shift conductivity is independent of $\omega$, as shown in Fig. \ref{fig:tpfall} (c,d).

\begin{figure}[]
\includegraphics[width=0.48\textwidth]{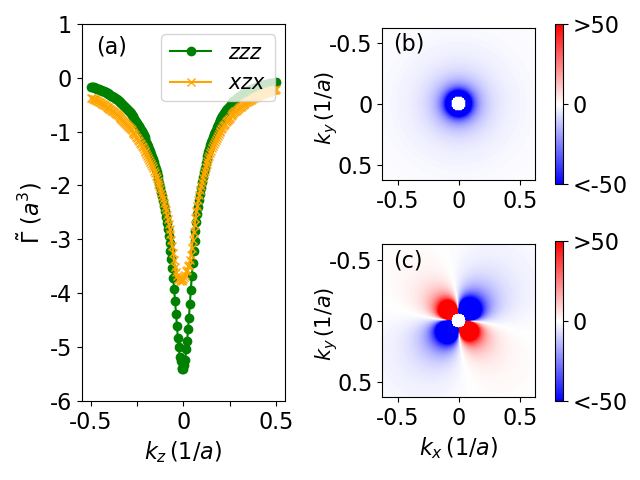}
\caption{Momentum resolved symplectic Christoffel symbol for TPF. (a) Along $k_z$, (b) on the $k_z=0$ plane for the $zzz$ component,
(c) on the $k_z=0$ plane for the $xzx$ component. The chemical potential is set slightly below the node,  $\mu=W_{TPF}-0.1$ eV.
}
\label{fig:christotpfxy}
\end{figure}

The relevant momentum resolved symplectic Christoffel symbols for linear shift conductivity $zzz$ and $zxx$ are shown in Fig. \ref{fig:christotpfxy}. The $zzz$ and $xzx$ components both show a peak near the node in the $k_z$ resolved plot. 
On the $k_z=0$ plane, the $zzz$ component is circularly symmetric, while the $xzx$ component shows mirror symmetry about the $k_x=k_y$ plane. 

\begin{figure}[b]
\includegraphics[width=0.48\textwidth]{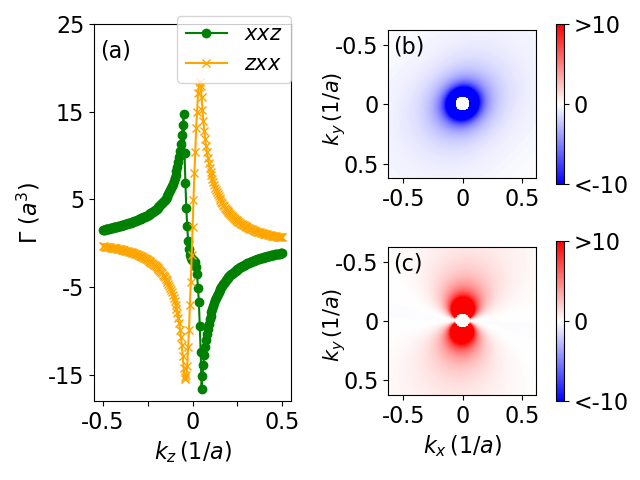}
\caption{Momentum resolved Christoffel symbol of the first kind ($\Gamma^{xxz}$ and $\Gamma^{zxx}$) for the TPF model.
(a) Along $k_z$, (b) on the $k_z=0$ plane for the $zzz$ component, (c) on the $k_z=0$ plane for the $xzx$ component.
The chemical potential is set slightly below the node,  $\mu=W_{TPF}-0.1$ eV.
}
\label{fig:christotpfcircxy}
\end{figure}

\begin{figure}
	\includegraphics[width=0.48\textwidth]{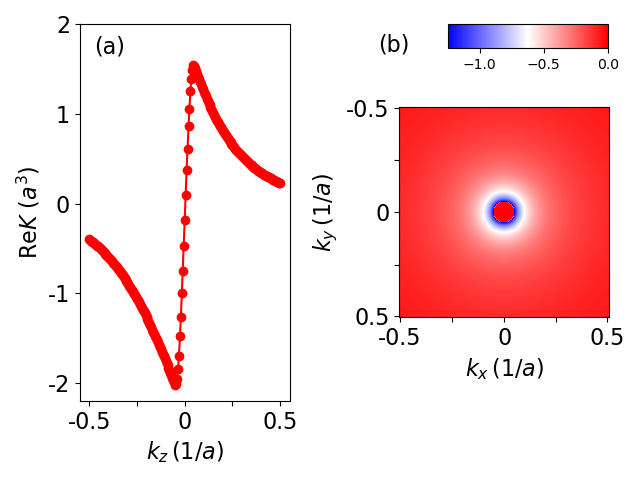}
	\caption{Momentum resolved contorsion tensor $Re\left[K^{xzy}\right]$ for the TPF model. This tensor contributes to the circular shift conductivity $zyx$. (a) Along $k_z$, (b) on the $k_z=0$ plane.}
	\label{fig:contorsion}
\end{figure}

\begin{figure}
\includegraphics[width=0.49\textwidth]{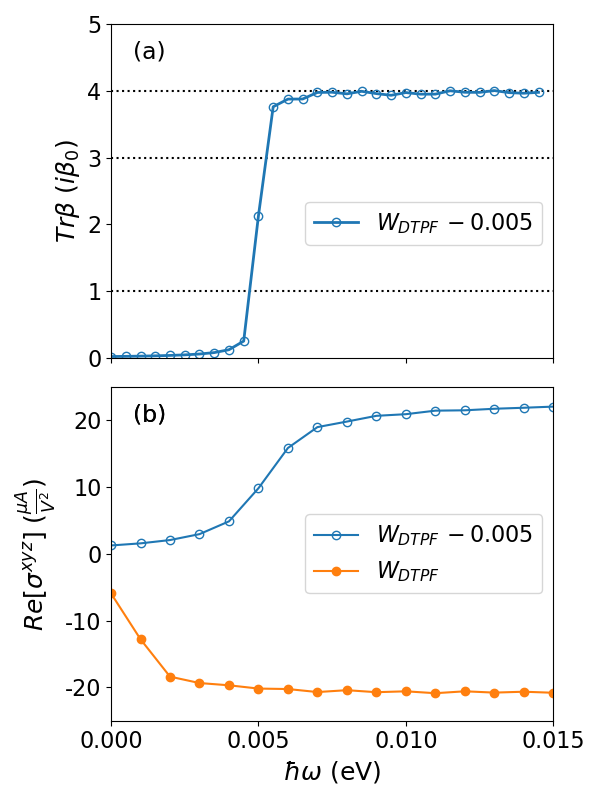}
\caption{(a) Circular injection and (b) linear shift conductivity for $H_{\Gamma198}$ without spin-orbit coupling
with chemical potential set slightly below the DTPF node ($\mu=-0.075$ eV $= W_{DTPF}-0.005$ eV).
In (b), the curve for chemical potential at the DTPF node
is also displayed.}
\label{fig:cosioffsocall}
\end{figure}

{\it Circular shift current.} Components of the circular shift conductivity are shown in Fig. \ref{fig:tpfall}(d) for $s=1$ and $b=1$. 
Interestingly, both $xzx$ and $zyx$ components of the circular shift conductivity vanish when $s=0$, showing that this term 
is vulnerable to the deformation of the on-site hopping. The corresponding Christoffel symbol of the first kind for  $\sigma^{xzx}$ 
is shown in Fig. \ref{fig:christotpfcircxy}. The $k_z$ resolved $\Gamma$ in Fig. \ref{fig:christotpfcircxy} (a) indicate 
that both $xxz$ and $zxx$ components change sign near the node $k_z=0$. 
For the circular shift $zyx$ component, the Christoffel symbols are zero along $k_z$. 
Equation \ref{eq:circshch} shows that the contribution to the $zyx$ circular shift conductivity is the contorsion tensor, 
not the Christoffel symbol. { Fig. \ref{fig:contorsion} shows the contorsion tensor $Re\left[K^{xzy}\right]$ in the momentum space. Because $Re\left[K^{xzy}\right]=-Re\left[K^{yzx}\right]$, only the $xzy$ component is shown. For other components of conductivities, the numerical values of the contorsion tensor is negligible compared to the Christoffel symbols; thus their contorsion tensors are not shown.}

If we further simplify the Hamiltonian to the linear order of $k$, the Berry curvature for each band of the TPF 
has a simple form $\mp \sin\theta, 0$ for the valence, conduction and flat band, respectively, 
thereby giving rise to the Chern number of $\mp 2, 0$. Nevertheless, the shift current vanishes after integration,
since the integrands are either 0 or odd functions. 

\subsection{Multifold fermions in the CoSi family}
In this subsection, we present the numerical results of the model Hamiltonians for multifold fermions in the CoSi family.
Since these model Hamiltonians have the time-reversal symmetry, only the linear shift current and circular injection current
would occur~\cite{Ahn2020}.

{\it Double triple point fermions.}  
$H_{\Gamma198}$ without SOC is a degenerate TPF, dubbed as double TPF (DTPF).
An symmetry analysis indicates that the $xyz$ element is the only nonvanishing independent component 
for both circular injection and linear shift current, which is plotted as a function of photon energy ($\hbar \omega$)
in Fig. \ref{fig:cosioffsocall}(a) and Fig. \ref{fig:cosioffsocall}(b), respectively.
Figure \ref{fig:cosioffsocall}(a) shows that for chemical potential being set slightly below the DTPF node ($\mu = W_{DTPF} - 0.005$ eV), 
the circular injection current is nearly zero when $\hbar \omega$ is smaller than the energy difference between $\mu$ and
$W_{DTPF}$ (0.005 eV). Nevertheless, it increases sharply when $\hbar \omega$ approaches to $0.005$ eV and quickly becomes
saturated as $\hbar \omega$ further increases. As a result of the topological charge carried by the degenerate point, 
the circular injection response would show quantization.
Figure \ref{fig:cosioffsocall}(a) indicates that the circular injection conductivity is quantized at 4 
when $\hbar \omega > 0.005$ eV because of the double degeneracy of the Weyl point with chiral charge 2. 

\begin{figure}[]
\includegraphics[width=0.48\textwidth]{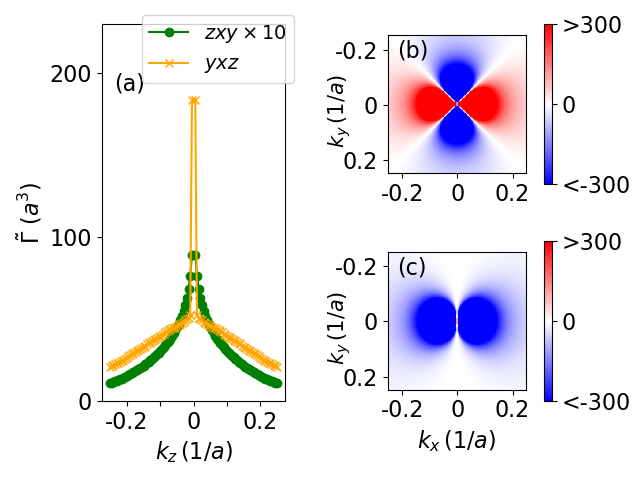}
\caption{Momentum resolved symplectic Christoffel symbol for the DTPF node at $\mu=W_{DTPF}$. 
(a) Along $k_z$, (b) on the $k_z=0$ plane for the $zxy$ component and (c) on the $k_z=0$ plane for the $yxz$ component. 
At the $\Gamma$ point, both $zxy$ and $yxz$ components diverge negatively, and thus, after the integration of $k_x,k_y$, the value is negative, as shown by the peak at $k_z=0$ in (a).  
        }
\label{fig:christodtpfxy}
\end{figure}

In Fig. \ref{fig:cosioffsocall} (b), the linear shift conductivity is shown. 
Interestingly, the shift conductivity for the chemical potential at the node and below the node are opposite in sign.  
{The major contribution comes from the quantum geometry of the flat band.  At low photon frequency, the flat band changes from being unoccupied at $\mu=W_{DTPF}-0.005$ to occupied at $\mu=W_{DTPF}$. This change is approximately equivalent to taking the complex conjugate of the Hermitian connection. Since the linear shift conductivity is given by the imaginary part of the Hermitian connection, the shift of chemical potential leads to the sign change. 
This is similar to the sign change of the Berry curvature when chemical potential shifts across the node. }
In both cases, the magnitude of the shift conductivity increases monotonically as $\hbar \omega$ increases from zero.
Similar to the circular injection current, the shift conductivity becomes saturated when $\hbar \omega$ is well above $0.005$ eV.
Nevertheless, in contrast to the circular injection current, the saturation of the shift conductivity apparently
does not result from its quantization behavior. {The saturation can be understood from power counting analysis, which shows that the lowest order of the shift conductivity is proportional to $a_{0}v'/v$.}

Since the linear shift conductivity comes from the divergent behavior of the symplectic Christoffel symbols 
near the topological nodes, we show in Figure \ref{fig:christodtpfxy} the symplectic Christoffel symbols for the DTPF node.
Figure \ref{fig:christodtpfxy} indicates that the $yxz$ component is more than one order of magnitude 
stronger than the $zxy$ component, the linear shift conductivity reveals mainly the $yxz$ component
of the symplectic Christoffel symbols. 

\begin{figure}
\includegraphics[width=0.49\textwidth]{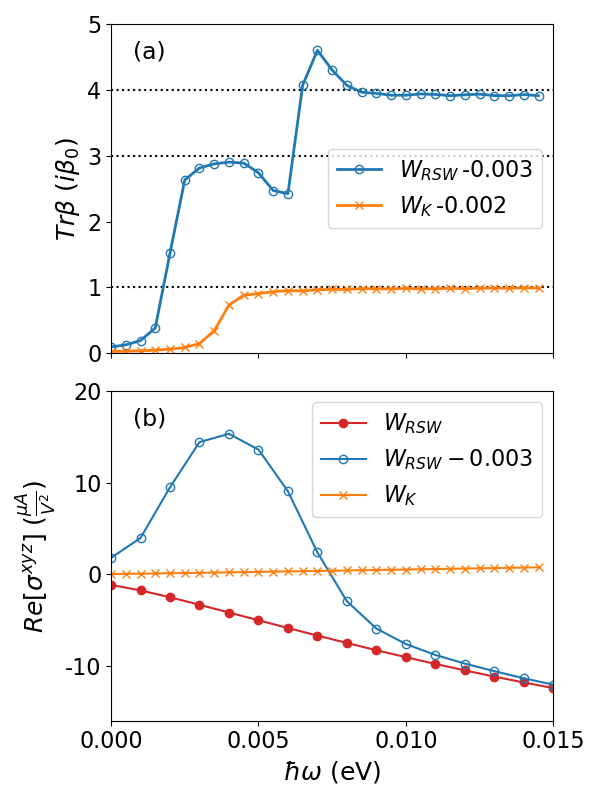}
\caption{(a) Circular injection  and (b) linear shift conductivity for $H_{\Gamma198}$ with spin-orbit coupling
(i.e., the RSW and Kramers nodes).
}
\label{fig:cosiall}
\end{figure}

{\it RSW fermions.} When the spin-orbit coupling is included in $H_{\Gamma198}$, the DTPF nodal point  
[see Fig. \ref{fig:198disp}(a)] splits into the RSW and Kramers nodes [see Fig. \ref{fig:198disp}(b)]. 
The calculated photoconductivity spectra for the RSW node are displayed in Fig. \ref{fig:cosiall}.
Figure \ref{fig:cosiall}(a) shows that the circular injection conductivity for the RSW fermions 
increases when $\hbar \omega$ approaches to 0.002 eV and becomes nearly saturated at $\sim$3 $\beta_0$
between $0.0025$ and $0.005$ eV. As $\hbar \omega$ further increases, it first dips slightly 
and then increases rapidly to the saturated value of 4 [see Fig. \ref{fig:cosiall}(a)]. 
This interesting behavior of the circular injection conductivity for the RSW node can be understood by the band dispersion of the RSW Hamiltonian 
displayed in Fig. \ref{fig:198disp}(b) where the RSW bands of RSW are labeled with blue numbers 1-4. 
When only the transition from the lowest band is active, the circular injection conductivity reveals 
the Chern number of the lowest band, which is $3$, and this explains the first plateau of $\sim$3. 
At higher photon frequencies, the transition between 
the second and the third band also occurs, giving rise to a quantization of $1$. The saturated value 
of the circular injection conductivity thus reveals the sum of the Chern numbers of the lowest two bands, which is $4$. 

The linear shift conductivity for the RSW node is displayed in Fig. \ref{fig:cosiall} (b). 
Interestingly, the conductivity in the low light frequency region below $\sim$0.007 eV changes sign 
when chemical potential is slightly lowered from the RSW node to $W_{RSW}-0.003$ eV. 
Specifically, when $\mu=W_{RSW}$ (red curve), the linear shift conductivity is negatively proportional to $\omega$. 
When $\mu=W_{RSW}-0.003$ eV (blue curve) (i.e. slightly below the RSW node), the conductivity shows a pronounced positive peak 
at the low frequencies. 
In this low frequency region, it can be seen from the band structure [Fig. \ref{fig:198disp}(b)] 
that the optically active bands are the second and third (first and second) for $\mu=W_{RSW}\ (W_{RSW}-0.003)$ eV. {
Thus, the linear shift conductivity reveals that the symplectic Chirstoffel symbols are opposite in sign between different pairs of bands.}
\begin{figure}[b]
\includegraphics[width=0.48\textwidth]{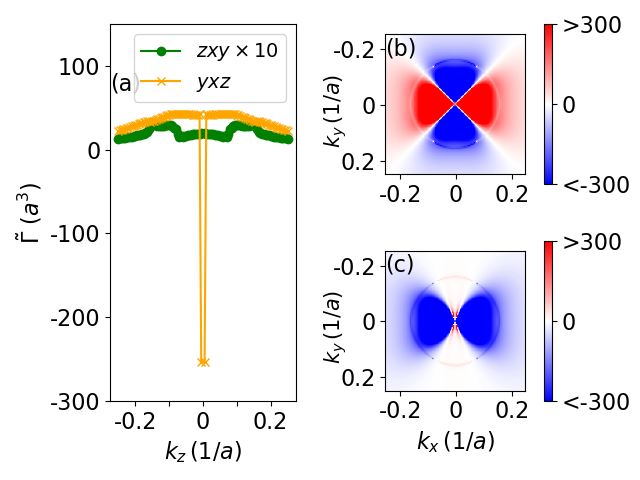}
\caption{Momentum resolved  symplectic Christoffel symbol for the RSW node, $\mu=W_{RSW}$. 
(a) Along $k_z$, (b) on the $k_z=0$ plane for $zxy$ component,
(c) on the $k_z=0$ plane for the $yxz$ component.
}
\label{fig:christorswxy}
\end{figure}
Moreover, for the $xyz$ componenet of the linear shift conductivity, the related components of the Christoffel symbols are $zxy$ and $yxz$. 
We calculate both components of the symplectic Christoffel symbols for the RSW fermions, as shown in Fig. \ref{fig:christorswxy}.
$\tilde{\Gamma}^{yxz}$ shows a very strong peak near $k_z=0$.
As for the DTPF node (Figure \ref{fig:christodtpfxy}), in contrast, $\tilde{\Gamma}^{zxy}$ is much weaker and no resonance is found at $k_z=0$. 
Thus, the major contribution to the linear shift conductivity 
is the $yxz$ component of the symplectic Christoffel symbol. The distribution of $\tilde{\Gamma}$ on the $k_z=0$ plane is also shown 
in Fig. \ref{fig:christorswxy} (b,c). There is a drastic change near the node.

After turning on the spin-orbit coupling in $H_{\Gamma198}$, the band structure changes drastically. The flat band in DTPF no longer exists in RSW node. The existence and absence of the flat band would alter the quantum geometry. The difference can be observed in comparing the symplectic Christoffel symbols [Fig. \ref{fig:christodtpfxy} and Fig. \ref{fig:christorswxy}]. 
As a consequence, the linear shift conductivity would have different behaviors. 	
Comparing the linear shift conductivity [Fig. [\ref{fig:cosioffsocall}(b) and Fig. \ref{fig:cosiall}(b)], the dependence on the photon frequency changes to be linear. The difference is likely to be the result of the large Christoffel symbols of the flat band.

{\it Kramers Weyl fermions.} 
The calculated photoconductivity spectra for the Kramers Weyl fermions are also shown in Fig. \ref{fig:cosiall}.
In this case, $\mu=W_K=-0.131$ eV, and the circular injection current probes the Chern number of the Kramer Weyl node. 
Thus, the circular injection conductivity is quantized at $1$ for $\hbar\omega > 0.005$ eV.

\begin{figure}[]
\includegraphics[width=0.48\textwidth]{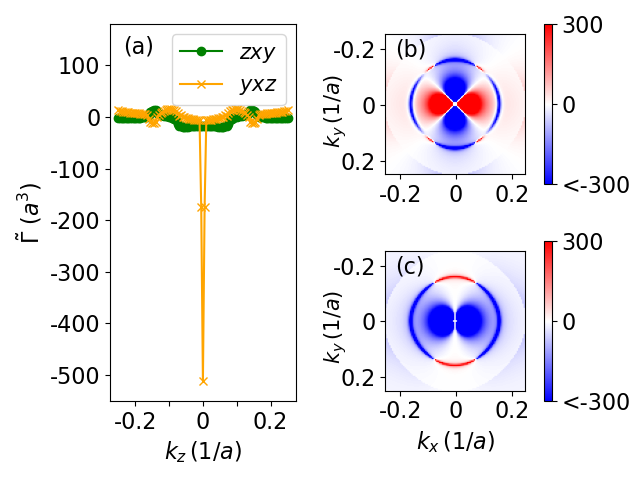}
\caption{Momentum resolved symplectic Christoffel symbol for Kramer Weyl of $H_{\Gamma198}$. $\mu=W_{K}\,-0.002$. (a) Along $k_z$, 
(b) on the $k_z=0$ plane for the $zxy$ component, (c) on the $k_z=0$ plane for the $yxz$ component. 
}
\label{fig:christokramerxy}
\end{figure}

Interestingly, the linear shift conductivity for the Kramers Weyl node is proportional to $\omega$, 
thus exhibiting the same trend as the type-I Weyl points \cite{Yang2018}.
In Fig. \ref{fig:christokramerxy}, the symplectic Christoffel symbols for the Kramer Weyl node are displayed,
which is the source of the linear shift current.
Figure \ref{fig:christokramerxy} thus indicates that the linear shift conductivity $\sigma^{x,yz}$ is dominated by the $yxz$ component 
of the symplectic Christoffel symbol, similar to that of the DTPF and RSW nodes shown above.

{The numerical results presented in this section were obtained under the assumption of zero absolute temperature in the Fermi-Dirac distribution.  
At finite temperature, the results for chemical potentials at the nodes would be different because the energy differences between bands are the smallest at the nodes and are more prone to thermal energies. For chemical potentials away from the nodes, as the smallest energy gap is larger than the thermal energy, the results would be qualitatively the same. 
}

{In addition, the numerical results are for low-energy effective Hamiltonian. The energy bands at much higher and lower energy regimes are ignored in the calculation. Thus, the conductivities are calculated at low photon frequency and valid for the energy regime where the nodes are isolated. }

\section{Discussion and Conclusion}\label{sec:concl}
The second-order photoconductivities and geometrical properties of chiral multifold fermions are studied in this paper. 
The analytical expressions for the injection and shift conductivities in terms of geometrical objects are given. 
As a result of the chiral symmetry breaking, the topological node and antinode are separated in energy. 
Thus, we study the second-order optical response of a single node. 
Our dimension analysis reveals that the lowest order of second-order photoconductivity is $\propto \omega^0$ 
and the second to the lowest order is $\propto \omega^1$. 
The quantities are calculated for the minimal symmorphic TPF model and the effective Hamiltonian for the CoSi family. 
Whether the $\omega^0$ term survives depends on the details of the Berry connections. 
For the TPF, RSW and Kramer Weyl nodes, the circular injection conductivity shows quantizations, as a result of the Chern number 
carried by the node. The linear shift conductivity for the RSW and Kramer Weyl node is $\propto \omega$. 
This behavior is similar to the type-I Weyl node. In contrast, the linear shift conductivity for the TPF node
is independent of $\omega$, but proportional to pseudo spin-orbit coupling. This relation has not been found 
in other Weyl semimetals, to the best of our knowledge. 
Furthermore, by analyzing the momentum-resolved geometrical objects, it is found that the quantum metric and Christoffel symbols 
are strongest near the nodes. The shift conductivities are related to contorsion tensors. The numerical results show that the contorsion tensors in general are at least one order of magnitude smaller than Christoffel symbols and symplectic Christoffel symbols for both model Hamiltonians. However, the contorsion tensors could be dominant. It is found that the circular shift conductivity $\sigma^{zyx}$ for the symmorphic TPF model is solely contributed by contorsion tensors, whereas the corresponding Christoffel symbols are zero. 
The study of these geometrical objects sheds light on the optical probe of the Hilbert space of lattices. 


 
\section*{acknowledgments}
H.-C.H., J.-S. Y. and G.-Y. G. acknowledge the support from the National Science and Technology Counsil (NSTC) and
the National Center for Theoretical Sciences (NCTS) in Taiwan. J.A. was supported by the Center for Advancement of Topological Semimetals, an Energy Frontier Research Center funded by the U.S. Department of Energy Office of Science, Office of Basic Energy Sciences, through the Ames Laboratory under contract No. DE-AC02-07CH11358.

\appendix
  
\section{Second order photoconductivities in terms of quantum geometrical quantities}\label{app:condgeom}
The second-order conductivity tensors are expressed in term of geometrical quantities in Eq. \ref{eq:linshch} and \ref{eq:circshch} of which the contorsion tensor  $K_{nm}^{bca}$ is defined as 
\begin{widetext}
\begin{eqnarray}
	K_{nm}^{bca}&=\frac{i}{2}\left[
	r_{nm}^b\sum\limits_{p\neq m,n}(r^c_{mp}r_{pn}^a-r^a_{mp}r^c_{pn})+
	r_{nm}^a\sum\limits_{p\neq m,n}(r^b_{mp}r_{pn}^c-r^c_{mp}r^b_{pn})
	-r_{nm}^a\sum\limits_{p\neq m,n}
	(r^a_{mp}r_{pn}^b-r^b_{mp}r^a_{pn})
	\right]\nonumber\\
	&-\frac{i}{3}{\rm Re}\left[r_{nm}^a\sum\limits_{p\neq m,n}(r^b_{mp}r_{pn}^c-r^c_{mp}r^b_{pn})
	-r_{nm}^c\sum\limits_{p\neq m,n}(r^a_{mp}r_{pn}^b-r^b_{mp}r^a_{pn})
	\right]+S_{bca}^{nm}.
\end{eqnarray}
\end{widetext}
Here, $S^{nm}_{bca}$ is imaginary and fully symmetric with respect to the permutation of $b,c$, and $a$. Since Eq. \ref{eq:chris} is satisfied with any choice of $S^{nm}_{bca}$, we take $S^{nm}_{bca}=0$ in this work.

\section{Symmetry analysis for the Berry curvature under $C_4$ symmetry}\label{app:sym}
The Berry curvature is the curl of the Berry connection $\mathbf{\Omega}_n=\nabla\times\mathbf{A}_n$, where $n$ is the band index and $\mathbf{A}_n=\langle n|i\nabla|n\rangle$. Under $C_4$ rotation symmetry, $k_x\rightarrow k_y, \ k_y\rightarrow -k_x, \  k_z\rightarrow k_z$. Because the Hamiltonian preserves $C_4$ symmetry, the Berry connection transforms as $A_n^x\rightarrow A_n^y, A_n^y\rightarrow -A_n^x,A_n^z\rightarrow A_n^z$. Thus, as required by the symmetry condition, the Berry curvature obeys
\begin{widetext}
\begin{eqnarray}
	\Omega_n^x(k_y,-k_x,k_z)&=&\Omega_n^y(k_x,k_y,k_z)\nonumber\\
	\Omega_n^y(k_y,-k_x,k_z)&=&-\Omega_n^x(k_x,k_y,k_z)\nonumber\\
	\Omega_n^z(k_y,-k_x,k_z)&=&\Omega_n^z(k_x,k_y,k_z)
\end{eqnarray} 
\end{widetext}

Since $C_4$ symmetry implies $C_2$ symmetry, the effect of $C_2$ is analyzed below. Under $C_2$ rotation symmetry, $k_x\rightarrow -k_x, \ k_y\rightarrow -k_y, \  k_z\rightarrow k_z$. The Berry connection transforms as 
$A_n^x\rightarrow A_n^y, A_n^y\rightarrow -A_n^x,A_n^z\rightarrow A_n^z$. Thus, the symmetry condition requires  
  \begin{widetext}
  	\begin{eqnarray}
  		\Omega_n^x(-k_x,-k_y,k_z)&=&-\Omega_n^x(k_x,k_y,k_z)\nonumber\\
  		\Omega_n^y(-k_x,-k_y,k_z)&=&-\Omega_n^y(k_x,k_y,k_z)\nonumber\\
  		\Omega_n^z(-k_x,-k_y,k_z)&=&\Omega_n^z(k_x,k_y,k_z).
  	\end{eqnarray} 
  \end{widetext}


\begin{thebibliography}{48}
\bibitem{foot:logr}{In general, $r_{mn}^a$ is complex and not a unit vector. It can be written in the form $r_{mn}^a=|r_{mn}^a|e^{i\phi_{mn}^a}$. Thus, $\log(r_{mn}^a)$ gives rise to two parts, $\log(r_{mn}^a)=\log(|r_{mn}^a|)+i\phi_{mn}^a$.}
\bibitem{foot:relaxapp}{A relaxation mechanism characterized by the relaxation time $\tau$ is assumed to obtain the steady-state injection current and photoconductivity. It is assumed that the relaxation time is independent of energy and momentum.}
\end{thebibliography}
\end{document}